\documentclass[a4paper]{jpconf}
\usepackage{graphicx}
\usepackage{amsmath}
\usepackage{amsfonts}
\usepackage{amssymb}

\begin{document}
\title{Fluxoid formation: size effects and non-equilibrium universality}

\author{David J. Weir and Ray J. Rivers}
\address{Blackett Laboratory, Imperial College\\
London SW7 2AZ, U.K.}
\ead{david.weir03@imperial.ac.uk, r.rivers@imperial.ac.uk}

\begin{abstract}
Simple causal arguments put forward by Kibble and Zurek suggest that the scaling behaviour of condensed matter at continuous transitions is related to the familiar universality classes of the systems at quasi-equilibrium. Although proposed 25 years ago or more, it is only in the last few years that it has been possible to devise experiments from which scaling exponents can be determined and in which this scenario can be tested.

In Ref.~\cite{ray1}, an unusually high Kibble-Zurek scaling exponent was reported for spontaneous fluxoid production in a single isolated superconducting Nb loop, albeit with low density. Using analytic approximations backed up by Langevin simulations, we argue that densities as small as these are too low to be attributable to scaling, and are conditioned by the small size of the loop. We also reflect on the physical differences between slow quenches and small rings, and derive some criteria for these differences, noting that recent work on slow quenches does not adequately explain the anomalous behaviour seen here.
\end{abstract}

\section{Overview}

Causality imposes strong constraints on the non-equilibrium behaviour of condensed matter systems, particularly in the vicinity of continuous transitions where, adiabatically, correlation lengths become arbitrarily large. Since, in reality, disorder-order transitions take place in a finite time, correlation lengths must remain finite; causality imposes a maximum speed at which the system can order itself. The greater the transition rate, the more constrained the correlation lengths will be. Should the ordered system be frustrated, this frustration will be relieved by the spontaneous creation of topological defects, whose separation will reflect this finiteness.

The suggestion that causality alone could lead to the spontaneous creation of topological defects was originally made by Kibble \cite{kibble1,kibble2}, not for condensed matter systems, but for the Grand Unified Theories (GUTs) present in the very early universe, which possess a cascade of transitions, almost all of which lead to the creation of cosmic strings (field vortices).   Zurek,  independently, had arrived at the same conclusions for the spontaneous formation of defects for condensed matter systems \cite{zurek1a,zurek1b,zurek2}, and this mechanism is known as the Kibble-Zurek (KZ) scenario. This has led to an active collaboration between cosmologists and condensed matter physicists over many years, most notably over the classification of defects and the similarities in the symmetry-breaking patterns and defects of liquid $^3He$ and GUTs \cite{volovik}. However, the more important aspect of the KZ scenario lies in the prediction of defect densities after a transition, since this should be amenable to experimental confirmation in condensed matter physics. Unfortunately, we cannot yet say the same for the defects of the early universe.

The simplest transitions are for systems cooled from one phase to another. Consider a temperature quench with quench time $\tau_Q = T_c(dT/dt)^{-1}_{T_c}$ in the vicinity of a continuous thermal transition with critical temperature $T_c$. It is predicted that, for a uniform quench applied to a homogeneous and isotropic system, the initial separation $\bar\xi$ between defects takes the scaling behaviour
\begin{equation}
 \bar\xi\approx \xi_0(\tau_Q/\tau_0)^{\sigma}
 \label{xibar}
\end{equation}
where $\xi_0$ and $\tau_0$ are fiducial distance and relaxation time scales, which depend upon the system. The important point is that the exponent $\sigma$ is derived from the adiabatic critical exponents and therefore shows universality.

There are several ways to derive (\ref{xibar}) from causal bounds \cite{kibble2,zurek2}. Most simply, we assume that the true correlation length tracks the adiabatic correlation length until the latter's rate of change is forced to try to exceed the relevant causal speed, whereupon the system freezes in, to unfreeze when the correlation length is again changing sufficiently slowly to be causally valid. This behaviour has been observed in a variety of simulations (e.g. see Refs. \cite{lagunab,lagunae}).

There have been many experimental tests of (\ref{xibar}) that give results commensurate with it without being capable of determining $\sigma$, although not necessarily for temperature quenches.  However, some experiments have been able to show scaling behaviour with quench time, in particular experiments on Rayleigh-Bernard convection patterns \cite{casado}, charge density waves in $TbTe_3$ \cite{yusupov} and fluxons in annular Josephson junctions \cite{monaco1,monaco2}. Several proposals for future experiments have been made, such as experiments for the formation of kink defects in linear ionic crystals \cite{delcampo} and quantum transitions in cold gases (e.g. Refs. \cite{zurek4,saito,dziarmaga}).

Although this was the original motive, attempts to see these results as reflections of transitions in the early universe have to be taken with extreme caution. The early universe is lacking in boundaries and homogeneous and isotropic, to which (\ref{xibar}) applies. Yet none of these attributes are possessed by laboratory condensed matter systems. In particular, inhomogeneity goes hand in hand with finite size. Inhomogeneity comes in two kinds. Firstly, that due directly to the finiteness of the system e.g. a trapped condensate has variable density \cite{zurek3}. The second is concerned with the inhomogeneous application of the quench. It is possible to incorporate such behaviours within scaling laws, but the end results are complicated \cite{zurek3,delcampo2}.

One way to try to avoid the worst of these problems is to use annular systems, originally proposed by Zurek \cite{zurek2}.  Such annuli are achievable in superconductors, Josephson junctions and in annular traps for cold bosonic gases \cite{dziarmaga}. If the circumference of the ring can be taken to be larger than its other dimensions for the purposes of domain formation, then the periodicity of the ring means we can have a homogeneous system without boundaries. Further, if the ring can be kept small, the problem of quench inhomogeneity becomes less pressing. The disadvantage of finding less than one defect, on average, is mitigated by ease of observation and the absence of defect annihilation. There is yet a further reason for choosing small rings of superconductors. The KZ mechanism is not the only reason for the spontaneous generation of magnetic flux in superconductors. The long wavelength modes of the magnetic field can freeze in spontaneously, almost without regard to the superconductor itself. This mechanism, the Hindmarsh-Rajantie \cite{rajantiea,rajantieb} mechanism gives contributions to the spontaneous flux which have a totally different behaviour from (\ref{xibar}).  The effect can be ignored for small superconductors. It was with all these in mind that the experiments on fluxon formation in annular Josephson junctions were performed \cite{monaco1,monaco2}, which showed remarkable allometric behaviour of the form given in (\ref{xibar}).

The anticipated scaling exponent for annular Josephson tunnel junctions was unclear, because of ambiguities in the fabrication that may or may not have allowed for proximity effects \cite{golubov}. However, single superconducting annuli should be less ambiguous and a recent experiment on low-temperature superconducting annuli  showed a higher than expected scaling exponent \cite{ray1}. This experiment is being revisited to look for possible systematic errors, but it is timely to see what effects due to size we might expect in superconducting systems that are smaller than the correlation length at the time of defect formation. We have some understanding of this for the quantum transitions of Bose gases \cite{saito,dziarmaga}, but these are controlled by the Gross-Pitaevskii equation whereas for superconductors phase ordering is associated with diffusion.

The purpose of this paper is to advance numerical and analytical arguments that the behaviour of an annular system changes from KZ scaling to exponential suppression for defect densities as small as those reported in Ref.~\cite{ray1}.  Such densities have not been widely studied in the literature, and certainly have not been widely explored for annular systems where the qualitative behaviour might be expected to be rather different from that for Bose-Einstein condensation experiments.

In Section \ref{sec:gaussian} we estimate the effect of small size by making a Gaussian approximation for a diffusive system on a one-dimensional ring. It suggests an interpolation between the canonical scaling behaviour for large rings and exponential damping. This is followed in Section \ref{sec:numerical} by a demonstration that this is confirmed by a Langevin simulation for a simple complex field on the ring. More realistic simulations are proposed in Section \ref{sec:environs}. Our conclusions can be found in Section \ref{sec:conclusions}.

\section{Gaussian approximation}
\label{sec:gaussian}

Although causality does impose bounds, the KZ argument as presented above is too simple.  It makes more sense to think of defect formation as due to the growth of long-wavelength unstable modes, where the growth of these is causally constrained. Surprisingly, treating the order parameter field as a Gaussian variable satisfying (short-time) linearised equations, which embodies this picture, recreates the scaling laws (\ref{xibar}) \cite{karra,bowick} and some key aspects of the nature of the transition \cite{moro}. It is borne out not only by the one-dimensional numerical simulations presented here, but also by more realistic simulations including the gauge field \cite{nextone}.

With due caution, we look to the effect of finite size on the winding number of a complex field $\Phi(x) = \chi(x)~e^{i\theta (x)}$ confined to a loop parameterised by the coordinate $x$. We see this as a proxy for a Cooper pair field confined to a small superconducting annulus of circumference $C$. On quenching the system from the normal to the superconducting phase, we can define a winding number density $n(x) = \partial_x\theta(x)/2\pi$,
where $0\leq x < C$ is the distance along the annulus.
The total winding number $n$ around the loop is zero on average, but will have non-zero variance $(\Delta n)^2=\langle n^2\rangle$, and it is this that we measure. The relevant quantity is the two-point correlation function
$\left\langle n\left(x\right)n\left(y\right)\right\rangle.$
We  assume that the complex order parameter field $\Phi = (\phi_1 + i\phi_2)/\sqrt{2}$
 is a Gaussian variable, whereby $\phi_1$ and $\phi_2$ are,
independently, Gaussian fields.

The correlation function for the winding number density is now determined by the correlation function $G(x)$ for the field components, defined by
\begin{eqnarray}
\left\langle\phi_{a}\left(x\right)\phi_{b}\left(y\right)\right\rangle
=\delta_{ab}G\left(\left|x-y\right|\right).
\label{eq:twentyfour}
\end{eqnarray}
A simple approximation gives
\begin{eqnarray}
\left\langle n\left(x\right) n\left(y\right)\right\rangle
&\approx&\frac{2}{(2\pi)^2}\left[f'\left(x\right)^{2}-f''\left(x\right)f\left(x\right)\right] %
\label{eq:fortyfive}
\end{eqnarray}
where
$f\left(x\right) = G(x)/G(0)$ and
$\left\langle\phi_{a}\left(x\right)\phi_{b}\left(y\right)\right\rangle
=\delta_{ab}G\left(\left|x-y\right|\right)$.

For strongly coupled dissipative systems the linearised time-dependent Landau-Ginzburg equations suggest that,  for {\it long rings} ($C/{\bar\xi}\gg 1$) $f\left(r\right)$ can be approximated well by the Gaussian form $f\left(r\right)\approx \exp(-r^{2}/2\xi_r^{2})$
where $\xi_r\approx {\bar\xi}$. Although this expression does not show periodicity, for long rings this should not matter. Substituting in (\ref{eq:fortyfive}) gives $\langle n^2\rangle = {\cal{O}}(C/\xi_r)$
in the case where $C$ is very large and the periodicity in the
field can be ignored. This linear dependence of $\langle n^2\rangle$ on $C$ is just as we
would expect from a random walk in phase along the ring and is assumed as input in the KZ picture.
Linearising the equations of motion for fast quenches then gives (\ref{xibar}) with $\sigma = 1/4$, in agreement with the KZ prediction.

 On the other hand, for short rings ($C/{\bar\xi}\ll 1$) we must have $f(x)$ periodic in $x$ (mod $C$).  To implement this we need to replace $f(x)$  by its Jacobi $\vartheta$ function generalisation
\begin{eqnarray}
 f(x)_{per} = \frac{\vartheta_3(\pi x/C|2\pi i\xi_r^2/C^2)}{\vartheta_3(0|2\pi i\xi_r^2/C^2)}
 \approx& 1- 4\sin^2(\pi x/C)~e^{-2\pi^2\xi_r^2/C^2},
\end{eqnarray}
whence $\langle n^2 \rangle = {\cal O}(e^{-4\pi^2\xi_r^2/C^2})$.
Thus, rather than the power falloff for large loops we have exponential damping
\begin{eqnarray}
\label{eq:expdamping}
\ln \langle n^2 \rangle &\approx& -4\pi^2\xi_r^2/C^2 + {\mbox const.}
\label{smallandslow}
\end{eqnarray}
This is in contrast to the proposal made in \cite{ray1} that the slope be doubled, which is more likely for a disc.
We should not push this approximation too far but it suggests exponential damping in the quench time whenever $\langle n^2\rangle \ll 1$, whether this be for very slow quenches in large rings or faster quenches in small rings.

\section{Numerical simulations}
\label{sec:numerical}

Our Langevin simulations are for the $\mathrm{U}(1)$ scalar field theory described above in $1+1$ dimensions with periodic boundary conditions, on a ring of circumference $C$. The potential is
\begin{equation}
\label{eq:pot}
V(|\Phi|^2) = \frac{1}{2} a(t) |\Phi|^2 + \frac{1}{4} |\Phi|^4,
\end{equation}
where  $a(t)$ is decreased linearly in time as $a(t) = - t/\tau_Q~~(t< \tau_Q)$ and $a(t) = -1~~(t>\tau_Q)$ to model a slow quench.
We start at $t=-2\tau_Q$ and continue until $t=4\tau_Q$, by which time the defects have frozen out.

To study this time-dependent system with dissipation and coupled to a heat bath, we consider the second-order Langevin equation
\begin{equation}
\label{eq:langeq}
\partial_t^2 \phi_a - \partial_x^2 \phi_a + \eta \partial_t \phi_a + \frac{\partial V}{\partial \phi_a} = \zeta_a,
\end{equation}
with Gaussian noise
\begin{equation}
\left< \zeta_a(x',t') \zeta_b(x,t) \right> = 2 \eta T \delta(x'-x) \delta(t'-t) \delta_{ab}, \qquad \left< \zeta_a(x,t) \right>=0.
\end{equation}
We take $\eta=1$ (so that the system is heavily damped) and $T=0.01$. It can be shown that such a system satisfies a fluctuation-dissipation theorem. We use a stochastic leapfrog method to evolve the equations of motion consistently~\cite{Borrill:1996uq}.

In this model, there is no dynamical gauge field, a reasonable approximation for the present system; we are interested in small rings or slow quenches when relatively few defects form so $\langle n^2 \rangle \ll 1$.
Recent work \cite{cugliandolo} suggests that the KZ scaling result (\ref{xibar}) should be modified for long times after slow quenches where defect annihilation is of importance. However, as we are dealing with flux trapping at very low densities, annihilation dynamics should be unimportant.

Applying periodic boundary conditions, we the phase angle is obtained through $\theta(x) = \mathrm{Pr}\,\mathrm{arg}(\phi_2(x) + i\phi_1(x))$, with a discretised winding number density
$n(x) = (\theta(x+\delta x)-\theta(x))/\delta x,\; \mathrm{mod}\;2\pi$.
The main bias in statistics of the winding number obtained from simulations will be a slight undercounting due to discretisation effects. One can account for this by checking that the results are robust to changes in lattice spacing.

\begin{figure}
\begin{minipage}[b]{0.48\linewidth}
\centering
\includegraphics[scale=0.3,angle=-90,clip=true,trim=0mm 0mm 0mm 0mm]
{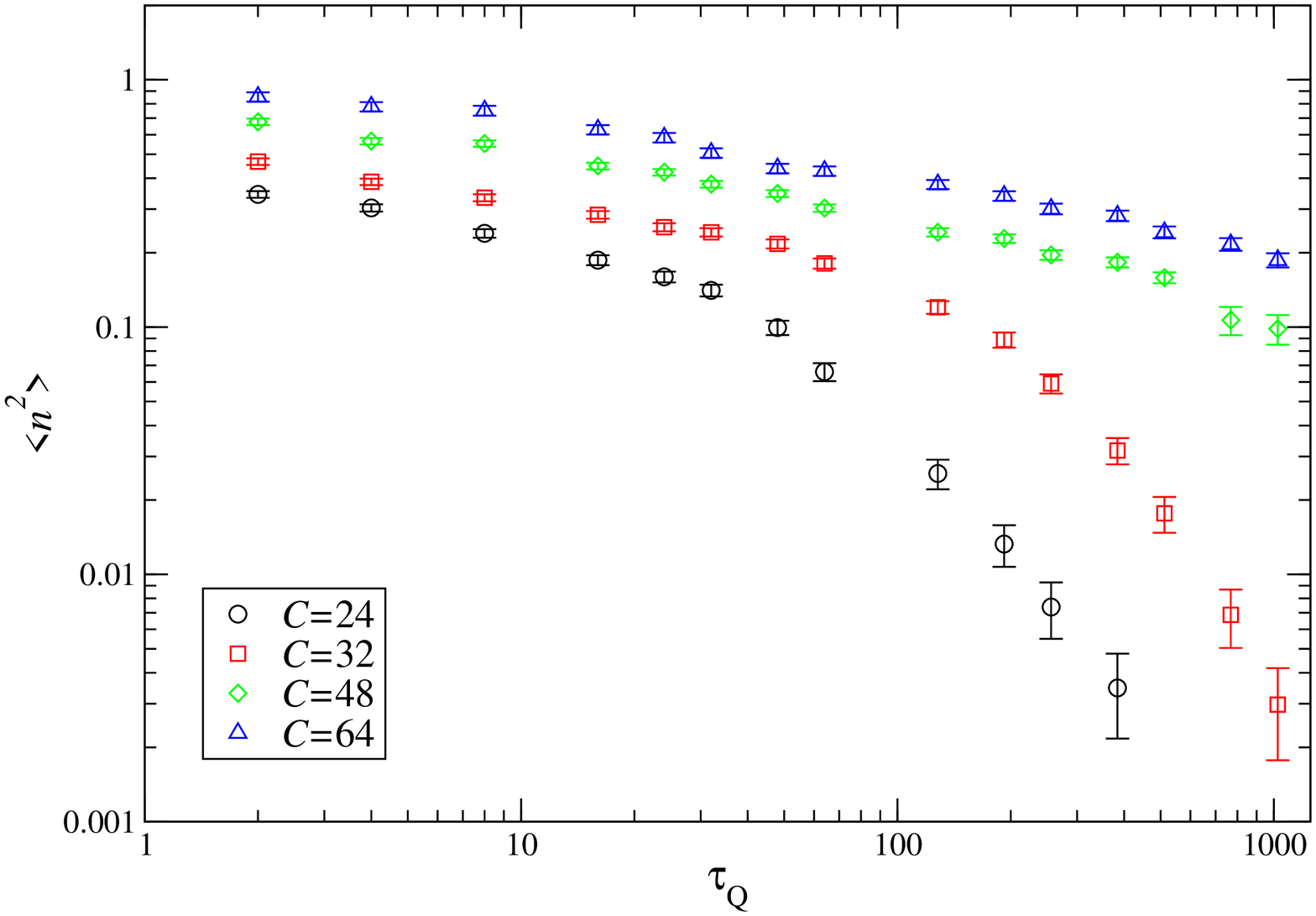}
\end{minipage}
\hspace{0.5cm}
\begin{minipage}[b]{0.48\linewidth}
\centering
\includegraphics[scale=0.3,angle=-90,clip=true,trim=0mm 0mm 0mm 0mm]{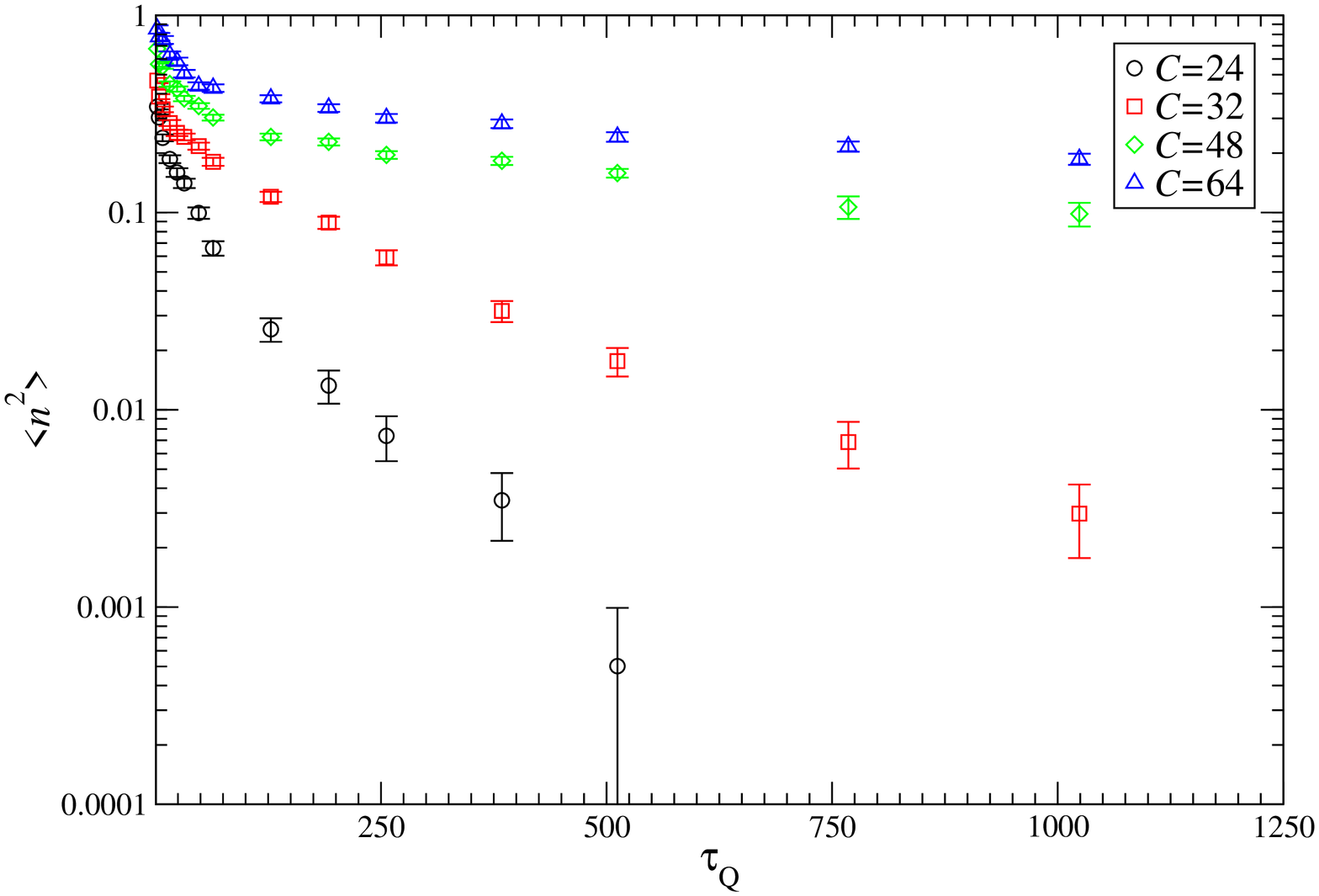}
\end{minipage}
\caption{\label{fig:volexpsup} Results of simulations, showing change between KZ and exponential suppression regimes. The doubly logarithmic plot at left shows KZ scaling with $\sigma \approx 0.25$ giving way to an exponential suppression regime below $\langle n^2 \rangle \approx 0.2$. The same data are plotted again on log-linear axes in the plot at right.}
\end{figure}

Our numerical results are ensemble averages of 2000 separate simulations for each combination of parameters. Typically, the lattice spacing is $\delta x=0.125$ and the timestep $\delta t = 0.0125$. The numbers of lattice sites used is in the region of $192-512$, giving circumferences $C$ between $24$ and $256$ in appropriate units. Qualitative checks of the system's sensitivity to discretisation effects were carried out, and decreasing the lattice spacing or the timestep by a factor of 2 did not affect the results presented here. The Gaussian noise was generated using the `ranlux' generator with Box-M\"{u}ller transformations.
In simulating the system, we need to quench the system rather slowly (potentially leading to growth of errors from the discretisation of the timestep), while keeping it fairly small (motivating a reduction in the lattice spacing). We think that our choice of parameters offers a reasonable trade-off.

In Figure \ref{fig:volexpsup} we see behaviour similar to that believed to have been seen in experiment \cite{ray1}: the KZ scaling with $\sigma \approx 0.25$ gives way at small defect densities, $\langle n^2 \rangle \approx 0.2$, to exponential suppression. This is consistent with (\ref{eq:expdamping}). It is easy to see how, for a limited set of data in the region of $\langle n^2 \rangle \approx 0.2$, finite-volume effects might be mistaken for a doubling of the scaling exponent.

The behaviour we see in Figure \ref{fig:volexpsup} does not particularly support the modified scaling behaviour at late times that was recently proposed \cite{cugliandolo}. In any case, it be unlikely that we would see defect interactions substantially alter a texture-forming quench of this kind at such low values of $\langle n^2 \rangle$.

\begin{figure}
\begin{minipage}[b]{0.5\linewidth}
\centering
\includegraphics[scale=0.3,angle=-90,clip=true,trim=0mm 0mm 0mm 0mm]
{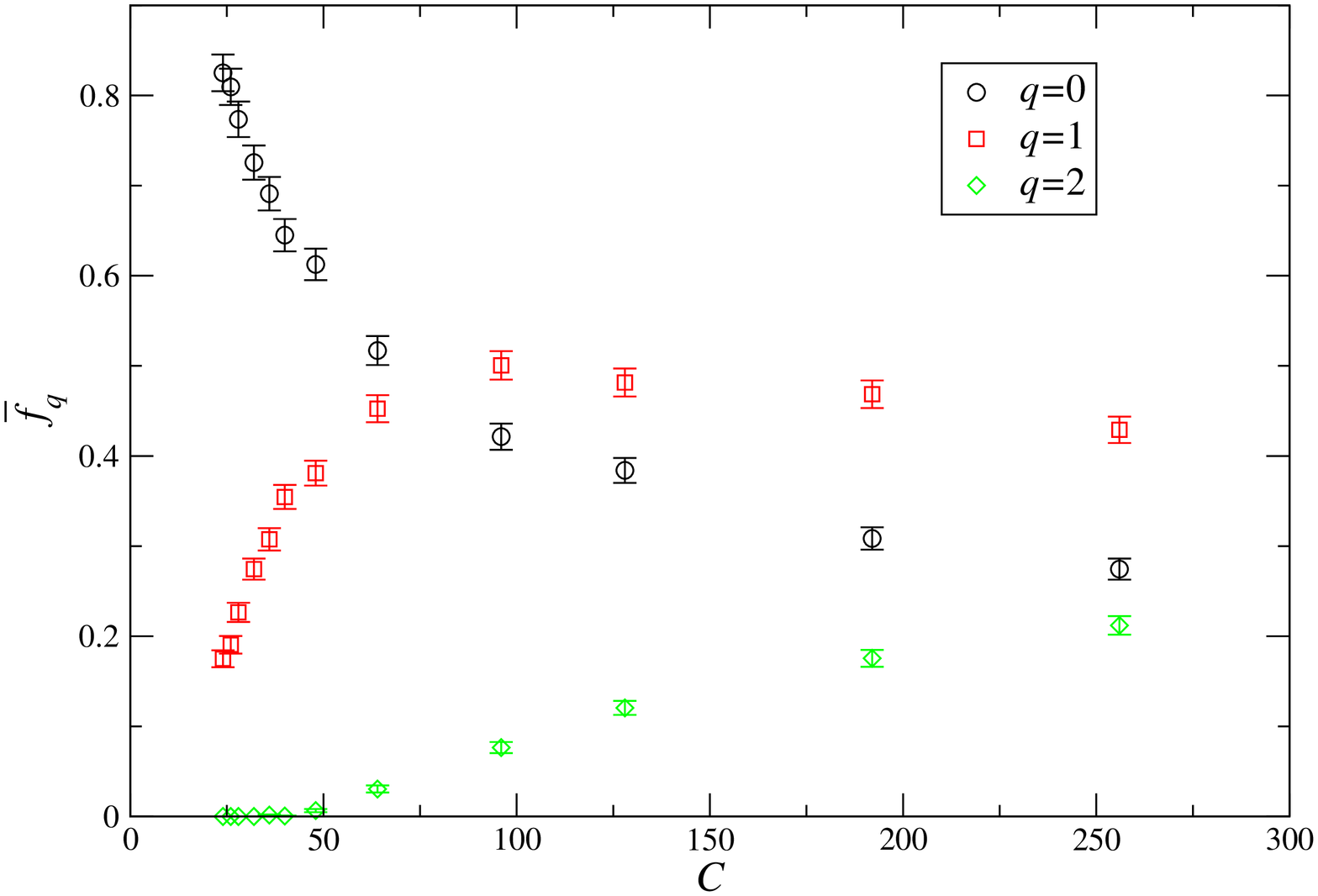}
\end{minipage}
\hspace{0.5cm}
\begin{minipage}[b]{0.5\linewidth}
\centering
\includegraphics[scale=0.3,angle=-90,clip=true,trim=0mm 0mm 0mm 0mm]{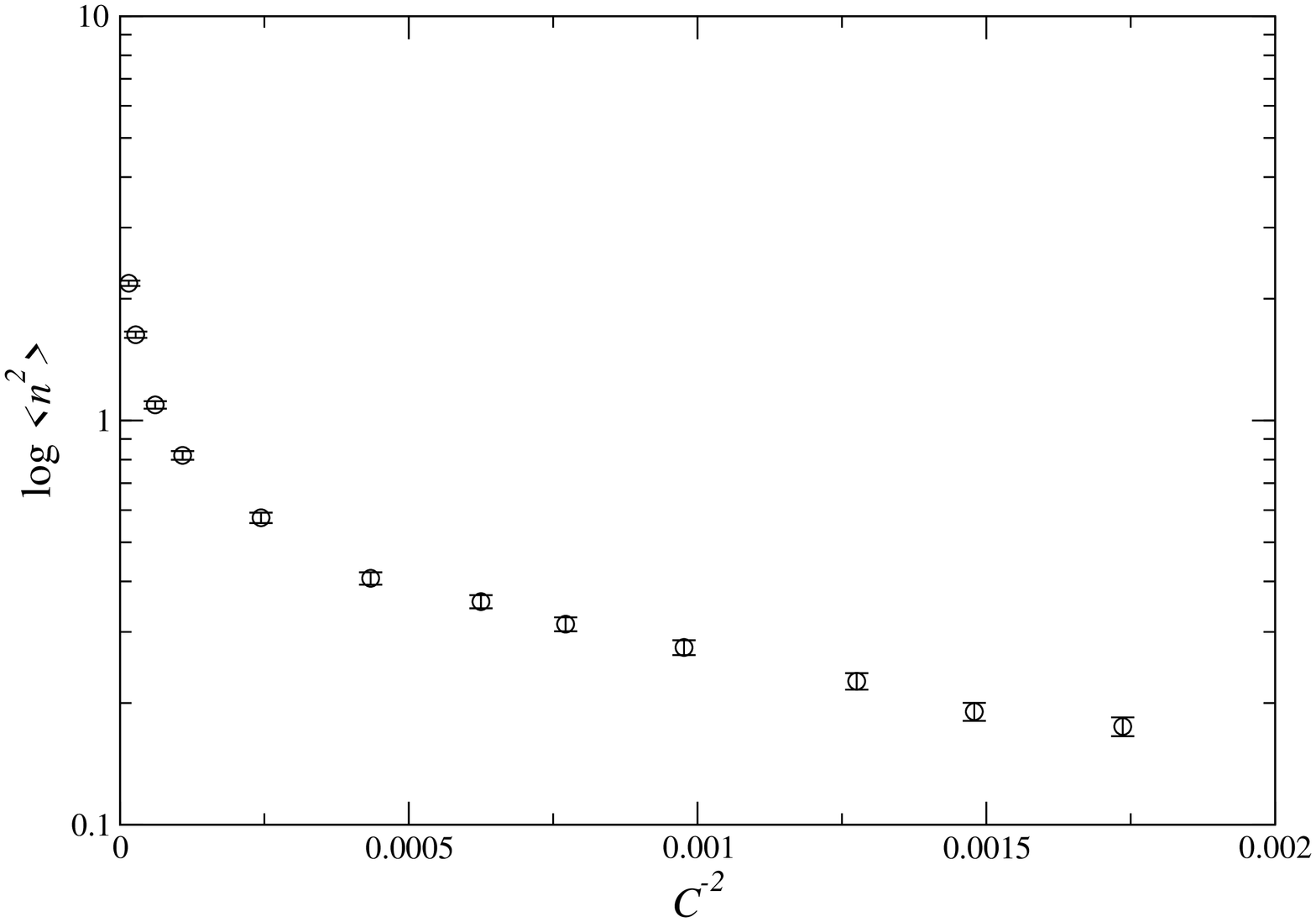}
\end{minipage}
\caption{\label{fig:histog2} At left is a plot of the probability $\bar{f}_q$ of finding a given number of fluxons with system size with $\tau_Q=24$; at right this has been combined into a measurement of $\langle n^2 \rangle$. The exponential scaling with $C^{-2}$ is clear for small volumes. Our simulations suggest that small volumes and slow quenches have the same effect.}
\end{figure}

While in experimental settings it is only possible to measure $\langle n^2 \rangle$, for the numerical simulation we can measure the probability of trapping exactly $q$ fluxons or antifluxons $\bar{f}_q$. At very small trapping probabilities where $\bar{f}_q \approx 0$ for $q\geq 2$, then $\langle n^2 \rangle \approx \bar{f}_1$. Figure \ref{fig:histog2} demonstrates the effect of system size on $\bar{f}_q$, and also on $\langle n^2 \rangle$. The results demonstrate that keeping the quench time relatively short and simply varying the circumference also supports the Gaussian approximation. In other words, slow quenches and small systems show the same exponential suppression of flux trapping, as follows from (\ref{smallandslow}).

\section{More realistic simulations: the system in its environment}
\label{sec:environs}

 More realistic simulations require the incorporation of a dynamical gauge field. It is not clear how valid the arguments advanced in the previous sections will be for the case of a thick ring, or where the system is still annular but is no longer circular. Genuine experiments also typically have stray magnetic fields that cannot be truly eliminated. None of these effects can be studied in $1+1$-dimensional simulation as there is no dynamical magnetic field. Simulations of the noncompact Abelian Higgs model in $2+1$ dimensions may be little better \cite{rajantiea}. For these reasons we favour treating the system as a thin ring in a larger simulational box containing only the $\mathrm{U}(1)$ field of vacuum electrodynamics \cite{nextone}.

There are no great technical obstacles to such a simulation. Performing a Langevin simulation while maintaining the Gauss constraint (having imposed temporal gauge $A_0 = 0$) is straightforward and has been used in previous Langevin simulations of similar systems, albeit with greater symmetry \cite{bettencourt}.

Indeed, the proposed simulation is not particularly elegant. In checking how external fields bias the trapping of flux, the penetration depth introduces a length scale which would be ignored for the simple case of a thin film.  It is neatest to use periodic boundary conditions, in which we leave a gap between the edge of the annulus and the sides of the simulational box; the net flux through the system must vanish and so trapped flux must return around the edges of the ring.

We can measure flux trapping in such a simulation by calculating the magnetic flux through the ring or by measuring the scalar field winding around the inner boundary of the system. The results from both methods are in agreement based on preliminary results \cite{nextone}.

Perhaps, if we wish to examine finite-size issues in effective models of planar (non-annular) superconducting films, the most elegant way may be to create a  film in the same way \cite{donaire}. Then a direct parallel can be drawn with the experiments carried out in Bose-Einstein condensates \cite{saito,dziarmaga}. Periodic boundary conditions may be employed without hesitation so long as a gap is left at the edges of the film.

\section{Conclusions}

\label{sec:conclusions}

 We have performed numerical simulations  for small annular systems with broken $U(1)$ symmetry, an idealisation of superconducting rings. We have demonstrated that the simple power behaviour of (\ref{xibar}) breaks down when the probability of finding a single defect falls below about $0.2$, more or less independently of the microscopic properties of the model. As we anticipated in a simple Gaussian approximation the likelihood of finding defects is then seen to become exponentially suppressed with quench rate or system size. This behaviour is very different from that of defect production in small Bose-Einstein condensates.

We have also discussed a potentially more valid simulational protocol for more realistic (planar) annular superconductors, incorporating the three dimensional electromagnetic field (and an external applied field) in a consistent way, and explained some of the limitations of such an approach.

\ack

RJR thanks R. Monaco for helpful discussions and both RJR and DJW are grateful to J.~Mygind and A. Gordeeva at Danish Technical University for their kind hospitality and helpful discussions. DJW also thanks A. Rajantie for useful discussions. DJW is funded by the Science and Technology Facilities Council. This work made use of the Imperial College High Performance Computing Service.


\section*{References}

\begin{thebibliography}{99}

\bibitem{ray1} Monaco R, Mygind J, Rivers R J, Koshelets V P 2009 {\it Phys. Rev.} B {\bf 80} 180501

\bibitem{kibble1} Kibble T W B 1976 {\it J. Phys.} A {\bf 9} 1387

\bibitem{kibble2} Kibble T W B 1980 {\it Phys. Rep.} {\bf 67} 183




\bibitem{zurek1a} Zurek W H 1985 {\it Nature} {\bf 317} 505
\bibitem{zurek1b} Zurek W H 1993 {\it Acta Phys. Pol.} B {\bf 24} 1301

\bibitem{zurek2} Zurek W H 1996 {\it Phys. Rep.} {\bf 276} 177


\bibitem{volovik} Volovik G E 2001 {\it Phys. Rep.} {\bf 351} 195


\bibitem{lagunab} Laguna P and Zurek W H 1998 {\it Phys. Rev.} D {\bf 58} 085021
\bibitem{lagunae} Antunes N D, Bettencourt L M A and Zurek W H 1999 {\it Phys. Rev. Lett.} {\bf 82} 2824


\bibitem{casado}  Casado S, Gonz\'{a}lez-Vi\~{n}as W, Mancini H and Boccaletti S 2001 {\it Phys. Rev.} E {\bf 63} 057301

\bibitem{yusupov} Yusupov R, Mertelj T, Kabanov V V, Brazovskii S, Kusar P, Chu J, Fisher I R and Mihailovic D 2010 {\it Nature~Phys.} {\bf 6} 681

\bibitem{monaco1} Monaco R, Mygind J and Rivers R J 2003 {\it Phys. Rev.} B {\bf 67} 104506

\bibitem{monaco2} Monaco R, Mygind J, Aaroe M, Rivers R J and Koshelets V P 2006 {\it Phys. Rev. Lett.} {\bf 96} 180604

\bibitem{delcampo} del Campo A, De Chiara G, Morigi G, Plenio M B, Retzker A 2010 {\it Phys. Rev. Lett.} {\bf 105} 075701


\bibitem{zurek4} Damski B and Zurek W H 2007 {\it Phys. Rev. Lett.} {\bf 99} 130402

\bibitem{saito} Saito H, Kawaguchi Y and Ueda M 2007 {\it Phys. Rev.} A {\bf 76} 043613

\bibitem{dziarmaga} Dziarmaga J, Meisner J and Zurek W H 2008 {\it Phys. Rev. Lett.} {\bf 101} 115701

\bibitem{zurek3} Zurek W H 2009 {\it Phys. Rev. Lett.} {\bf 102} 105702

\bibitem{delcampo2} del Campo A, Retzker A and Plenio M B 2010 Heterogeneous Kibble-Zurek mechanism: vortex nucleation during Bose-Einstein condensation {\it Preprint} 1010.6190

\bibitem{rajantiea} Hindmarsh M and Rajantie A 2000 {\it Phys. Rev. Lett.} {\bf 85} 4660
\bibitem{rajantieb} Rajantie A 2001 {\it J. Low Temp. Phys.} {\bf 124} 5







\bibitem{golubov}  Golubov A A, Houwman E P, Gijsbertsen J G, Krasnov V M, Flokstra J and Rogalla H 1995 \\
{\it Phys.~Rev.}~B~{\bf~51}~1073




\bibitem{karra} Karra G and Rivers R J 1998 {\it Phys. Rev. Lett.} {\bf 81} 3707

\bibitem{bowick} Bowick M and Momen A 1998  {\it Phys. Rev.} D {\bf 58} 085014

\bibitem{moro} Moro E and Lythe G 1999 {\it Phys. Rev.} E {\bf 59} R1303

\bibitem{nextone} Rajantie A, Rivers R J and Weir D J {\it In preparation}

\bibitem{Borrill:1996uq} Borrill J and Gleiser M 1997 {\it Nucl. Phys.} B {\bf 483} 416


\bibitem{cugliandolo} Biroli G, Cugliandolo L F, Sicilia A 2010 {\it Phys. Rev.} E {\bf 81} 050101

\bibitem{bettencourt} Stephens G J, Bettencourt L M A and Zurek W H 2002 {\it Phys. Rev. Lett.} {\bf 88} 137004

\bibitem{donaire} Donaire M, Kibble T W B, Rajantie A 2007 {\it New J. Phys.} {\bf 9} 148

\end{thebibliography}
\end{document}